\documentclass[preprint,showpacs,preprintnumbers,amsmath,amssymb]{revtex4}

\usepackage{graphicx}

\newcommand{\ket}[1]{\left|#1\right\rangle}
\newcommand{\bra}[1]{\left\langle#1\right|}

\def\*#1{\mathbf{#1}}

\begin{document}

\title{Spontaneous emission spectrum of a two-level atom in a very high Q cavity}

\author{Alexia Auff\`eves$^{1}$}%
\author{Benjamin Besga$^{1}$}%
\author{Jean-Michel G\'erard$^{2}$}%
\author{Jean-Philippe Poizat$^{1}$}%
\affiliation{$^{1}$CEA/CNRS/UJF Joint team " Nanophysics and
semiconductors ",Institut N\'eel-CNRS, BP 166, 25, av. des Martyrs,
38042 Grenoble Cedex 9, France}

\affiliation{$^{2}$CEA/CNRS/UJF Joint team " Nanophysics and
semiconductors", CEA/INAC/SP2M, 17 rue des Martyrs, 38054 Grenoble,
France}

\email{alexia.auffeves@grenoble.cnrs.fr}

\date{\today}

\begin{abstract}
In this paper we consider an initially excited two-level system
coupled to a monomode cavity, and compute exact expressions for the
spectra spontaneously emitted by each system in the general case
where they have arbitrary linewidths and frequencies. Our method is
based on the fact that this problem has an easily solvable classical
counterpart, which provides a clear interpretation of the evidenced
phenomena. We show that if the cavity linewidth is much lower than
the atomic linewidth, photons are emitted at the cavity frequency,
even if the atom and the cavity are strongly detuned. We also study
the links between the spontaneous emission spectra and the
fluorescence spectra emitted when the atom is driven by a classical
field of tunable frequency in the low excitation limit.
\end{abstract}

\maketitle

\section{Introduction}
The spontaneous emission (SE) properties of a two-level system
strongly depend on its electromagnetic environment.
Purcell~\cite{Purcell} predicted first that the linewidth of an atom
placed in a resonant cavity may be increased, and inhibited if the
cavity is off resonance. When the atom-cavity coupling $g$ becomes
larger than the atomic and the cavity linewidths, respectively
denoted $\gamma$ and $\kappa$, SE becomes reversible, giving rise to
the well-known vacuum Rabi oscillation~\cite{Brune}. These
oscillations define the so-called strong coupling regime, whereas
the regime where SE is irreversible is called weak coupling regime.
The strong coupling regime has been reached in various systems,
ranging from Rydberg atoms in microwave cavities~\cite{Brune},
alcaline atoms in optical cavities~\cite{Rempe, Boozer} to
semiconducting devices~\cite{Reithmaier,
Yoshie,Peter,Hennessy,Press,Englund}, and opens the way to the
implementation of fundamental experiments in the field of quantum
information. Two other regimes emerge from the comparison between
the respective linewidths of the emitter $\gamma$ and of the cavity
$\kappa$, namely the good emitter regime (resp. the good cavity
regime), fulfilling $\gamma < \kappa$ (resp. $\kappa < \gamma$). Up
to now, most experiments have been conducted in the good emitter
regime. Nevertheless, recent technological developments have greatly
improved the quality factor Q of the cavities used in cavity quantum
electrodynamics (CQED) experiments, allowing to enter the less
explored regime where the cavity linewidth is of the same order of
magnitude than the atomic linewidth, or even much smaller. The
typical Q factor of the best optical cavities is
$10^8$~\cite{Boozer,Rempe}, which corresponds to the linewidth of
the used transition in cesium and in rubidium atoms. Q factors as
high as $10^{10}$ have been reached in the microwave
domain~\cite{Seb}, which is ten times better than the Q factor of
the transition between the considered Rydberg states. In the field
of solid-state optical microcavities, Q's above $10^8$ have been
achieved for silica microspheres~\cite{Sandoghar} or
microtoroids~\cite{Armani}, whereas impressive progress has been
recently witnessed for semiconductor based cavities such as
micropillars~\cite{Reitzenstein}, microdisks~\cite{Srivanasan} or
photonic crystal cavities~\cite{Weidner}. On the emitter side, a
radiative-lifetime limited emission linewidth has been observed for
single semiconductor quantum dots at low temperature under resonant
pumping conditions~\cite{Langbein}. This linewidth corresponds to a
typical Q of $10^6$ for standard InAs self-assembled QDs
($\tau\sim1$~ns, $\lambda\sim1\mu$m); even smaller values in the
$10^5-10^4$ range can be achieved for giant-oscillator-strength QDs
obtained in the InGaAs/GaAs~\cite{Reithmaier} or
GaAs/AlAs~\cite{Peter,Andreani} systems.

After the pioneering work of Purcell, the theoretical effort to
understand and model the spectra emitted by the atom-cavity system
was pursued in~\cite{Sanchez}, where the emission spectrum of a
Rydberg atom embedded in a cavity with no losses was computed, and
the existence of the vacuum Rabi doublet was first predicted. The
emission and absorption spectrum of an ideal two-level atom in a
cavity with losses was studied in~\cite{Agarwal}, and the
fluorescence and spontaneous emission spectra emitted by a lossy
atom coupled to a lossy cavity was computed
in~\cite{Carmichael,Rice}, both systems being on resonance. Note
that the problem of a two-level atom trapped in an external
potential and strongly coupled to a monomode cavity is also actively
studied. In particular, new and efficient cooling mechanisms have
been evidenced~\cite{Hechenblaikner, Morigi1} and the fluorescence
spectra emitted by the atom and the cavity during the cooling
process have been computed~\cite{Morigi2}.

In this paper, we compute exact expressions for the spectra
spontaneously emitted respectively by the atom and by the cavity, in
the most general case where the atom and the cavity have arbitrary
linewidths and frequencies in the low excitation limit. Our results
match all previous studies, providing them with a common theoretical
frame, and goes beyond as it allows to investigate new unexplored
regimes. Our method is based on the existence of an easily solvable
classical counterpart, and offers a new insight on previously
predicted phenomena like cavity induced
transparency~\cite{Alsing,Rice,bibi} and Rayleigh
scattering~\cite{serge}. To our knowledge, this is the first
extensive study on the emission properties of the atom-cavity system
considered as a whole and accounting for the conservation of the
excitation number and energy. Moreover, the spectra emitted
respectively by the atom and the cavity can be compared in the good
emitter regime and in the good cavity regime defined above. The
comparison is striking when the atom and the cavity are detuned. In
the first case, both the atom and the cavity emit photons at the
atomic frequency. On the contrary, in the second case, the spectra
are dramatically different. In particular, we show that a
significant fraction of photons are emitted at the cavity frequency.
This apparently puzzling feature was experimentally observed
by~\cite{Reithmaier, Yoshie, Peter, Hennessy, Press}, the present
paper suggests here a simple theoretical explanation.

The paper is organized as follows. We start in section II by
computing the spontaneous emission spectrum recorded if the atom is
initially excited and the cavity empty, whether one looks at the
atom or at the cavity. In section III we relate these relaxation
spectra to the spectra which could be obtained when one drives the
atom or the cavity with a classical field. In sections IV and V we
give a physical interpretation of the results, in the good emitter
regime and in the good cavity regime respectively.

\section{Computation of the spontaneous emission spectra}

\begin{figure}[h,t]c
\begin{center}
\includegraphics[height=10cm]{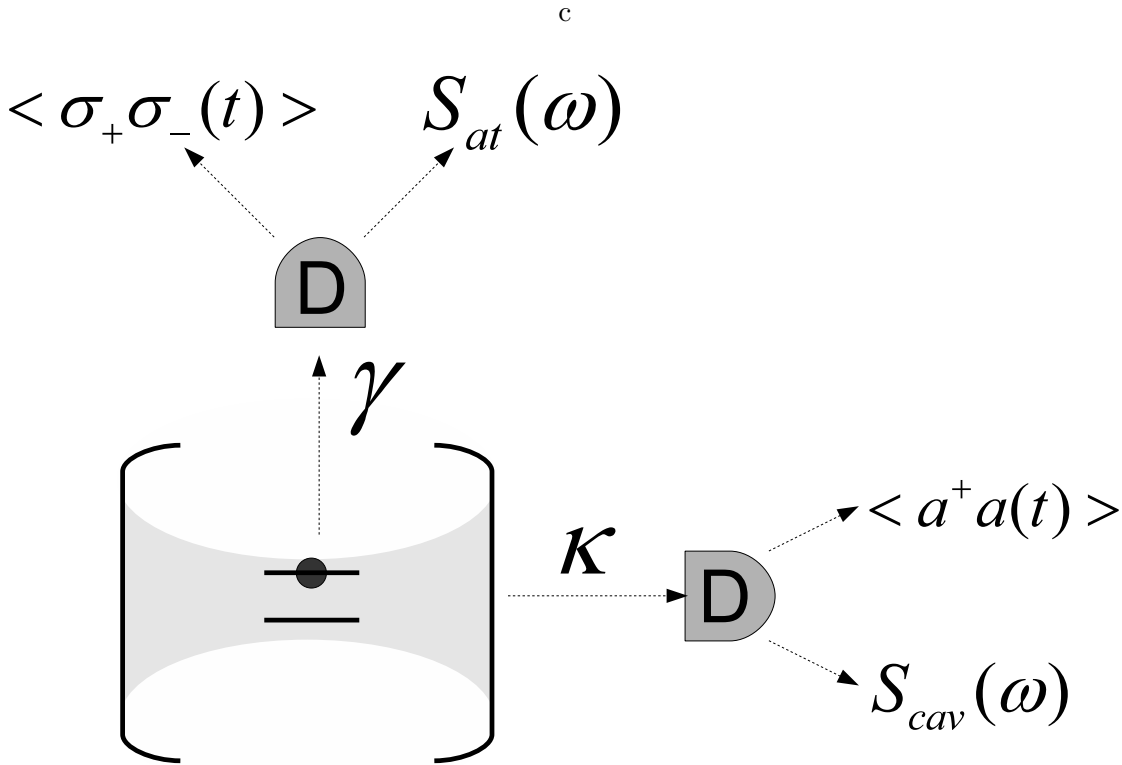}
\caption{\it System under study. An initially excited atom is
coupled to a empty cavity. Both systems have leaks characterized by
the relaxation rates $\gamma$ and $\kappa$. A time resolved /
frequency resolved detector $D$ is placed in the atomic (resp.
cavity) channel of losses, its probability to register a photon is
proportional to $\langle \sigma_+\sigma_-(t) \rangle$ (resp.
$\langle a^\dagger a(t) \rangle$) if it is time-resolved, and to
$S_{at}(\omega)$ (resp. $S_{cav}(\omega)$) if it is
frequency-resolved. } \label{fig:systeme}
\end{center}
\end{figure}

In this section, we detail the method to compute the spectra
spontaneously emitted by the atom, further denoted $S_{at}(\omega)$
and by the cavity, further denoted $S_{cav}(\omega)$, if the atom is
initially excited and the cavity is empty, as it is depicted in
figure~\ref{fig:systeme}. This initial state can be prepared in a
wide range of experimental systems. As an example, it corresponds to
the case of Rydberg atoms prepared in their excited state before
they enter the mode of a $0$~K microwave cavity~\cite{Seb}. It can
also be reached if the atom and the cavity are permanently coupled.
By non-resonantly pumping a quantum dot(QD) embedded in a
semiconducting cavity, one can feed the QD with a single exciton,
the cavity remaining empty. Coherent excitation of a Cooper box
capacitively coupled to a high Q transmission line has been
realized, allowing to generate single microwave photons with high
efficiency~\cite{Blais}.

To selectively register the spectrum $S_{at}(\omega)$ or
$S_{cav}(\omega)$, we connect a detector either to the atomic
channel or to the cavity channel of losses. From an experimental
point of view, such a distinction requires to use a cavity whose
emission diagram is directional. A predominant coupling with the
cavity channel of losses can be obtained by placing the detector
within the emission pattern, a better coupling with the atomic
channel being reached outside. As an example, the fundamental mode
of a micropillar is directional, as it emits photons in a small
solid angle. Consequently, it provides an efficient single photon
source~\cite{Moreau} and offers an interesting realization of
solid-state one-dimensional atom~\cite{bibi}. Moreover,
electromagnetic engineering provided by photonic crystal technology
allows to control the radiation diagram of the cavities and of the
leaky modes~\cite{Lee}.

We denote $\omega_0$ the atomic frequency, $\omega_{cav}$ the cavity
frequency, $\delta=\omega_{cav}-\omega_0$ the atom-cavity detuning.
The annihilation operator in the cavity mode is written $a$, and the
atomic operators $\sigma_-=\ket{g}\bra{e}$,
$\sigma_+=\ket{e}\bra{g}$ and
$\sigma_z=(\ket{e}\bra{e}-\ket{g}\bra{g})$ where $\ket{e}$ and
$\ket{g}$ are the excited and ground state of the atom respectively.
We take the quantity $\omega_M=(\omega_{cav}+\omega_0)/2$ as the
origin of frequencies. Energy can relax from the atom (resp. from
the cavity) into a continuum of empty modes at a rate $\gamma$
(resp. $\kappa$). We neglect any other dephasing processes. We can
write for the operators $a$ and $\sigma_-$ the following quantum
Langevin equations~\cite{Lax}:

\begin{equation}\label{equ:langevinq1}
\begin{array}{l}
{\displaystyle
\dot{a}=\left(-i\frac{\delta}{2}-\frac{\kappa}{2}\right)a+g\sigma_-+F_a
}
\\
{\displaystyle
\dot{\sigma}_-=\left(i\frac{\delta}{2}-\frac{\gamma}{2}\right)\sigma_-+g\sigma_za+F_s}.

\end{array}
\end{equation}

The functions $F_a$ and $F_s$ are the Langevin forces allowing the
conservation of the commutation relations. The spectra emitted by
the atom and by the cavity fulfill~\cite{Glauber}

\begin{equation}
\begin{array}{l}\label{glauber} {\displaystyle S_{at}(\omega)=\frac{\int dt
d\tau \theta(t)\theta(t+\tau)\langle \sigma_+(t+\tau)\sigma_-(t)
\rangle e^{-i\omega\tau}}{2\pi\int dt \langle
S_+(t)\sigma_-(t) \rangle}}\\
{\displaystyle S_{cav}(\omega)=\frac{\int dt d\tau
\theta(t)\theta(t+\tau)\langle a^\dagger(t+\tau)a(t) \rangle
e^{-i\omega\tau}}{2\pi\int dt \langle a^\dagger(t)a(t) \rangle}},\\
\end{array}
\end{equation}

which can be rewritten, evidencing their reality and showing the
positivity of the delay $\tau$

\begin{equation}
\begin{array}{l}\label{equ:phdet}
{\displaystyle S_{at}(\omega)=\frac{\int dt d\tau (\langle
\sigma_+(t+\tau)\sigma_-(t) \rangle
\theta(t)\theta(\tau)e^{-i\omega\tau} + cc)}{2\pi\int dt \langle
S_+(t)\sigma_-(t) \rangle}} \\
{\displaystyle S_{cav}(\omega)=\frac{\int dt d\tau (\langle
a^\dagger(t+\tau)a(t) \rangle \theta(t)\theta(\tau)e^{-i\omega\tau}
+ cc)}{2\pi\int dt \langle a^\dagger(t)a(t) \rangle}}.
\end{array}
\end{equation}

In the following, we show that the correlation functions $\langle
\sigma_+ (t+\tau) \sigma_-(t) \rangle$ and $\langle a^\dagger
(t+\tau) a(t) \rangle$ are identical to the correlation functions of
two coupled cavities ${\cal C}_1$ and ${\cal C}_2$ containing
classical fields whose amplitudes are respectively denoted
$\alpha(t)$ and $\beta(t)$, their complex frequencies being equal to
the atomic and cavity complex frequencies,
$\tilde{\omega}_1=\tilde{\omega}_{at}=-\delta/2-i\gamma/2$, \\
$\tilde{\omega}_2=\tilde{\omega}_{cav}=\delta/2-i\kappa/2$. The
cavity ${\cal C}_1$ has initially been fed ($|\alpha(0)|^2=1$) while
${\cal C}_2$ was empty ($|\beta(0)|^2=0$). This problem is solved in
appendix A.

We use the quantum regression theorem, which states that the
evolution of the quantities $\langle \sigma_+(t+\tau)X(t) \rangle$
and $\langle a^\dagger(t+\tau)X(t) \rangle$ (where $X(t)$ describes
either $a(t)$ or $\sigma_-(t)$) as a function of the delay $\tau>0$
obeys the same equation as the expectation values $\langle
\sigma_+(t) \rangle$ and $\langle a^\dagger(t) \rangle$. The initial
conditions, reached for $\tau=0$, correspond to the mean populations
$\langle \sigma_+(t)X(t) \rangle$ and $\langle a^\dagger(t) X(t)
\rangle$. As the average of the Langevin forces is $0$, these mean
values check the following equations

\begin{equation}\label{equ:langevinq1}
\begin{array}{l}
{\displaystyle \langle \frac{da}{dt}
\rangle=\left(-i\frac{\delta}{2}-\frac{\kappa}{2}\right)\langle a
\rangle +g \langle \sigma_- \rangle }
\\
{\displaystyle \langle \frac{d\sigma_-}{dt} \rangle
=\left(i\frac{\delta}{2}-\frac{\gamma}{2}\right)\langle \sigma_-
\rangle
+g \langle \sigma_za \rangle}\\

\end{array}
\end{equation}

and

\begin{equation}\label{equ:langevinq2}
\begin{array}{l}
{\displaystyle \frac{d \langle a^\dagger a \rangle }{dt}=-\kappa
\langle a^\dagger a \rangle + g
\langle \sigma_+ a \rangle + g \langle a^\dagger \sigma_- \rangle }\\
{\displaystyle \frac{ d \langle \sigma_+\sigma_- \rangle}
{dt}=-\gamma \langle \sigma_+\sigma_- \rangle - g \langle \sigma_+ a
\rangle - g
\langle a^\dagger \sigma_- \rangle }\\
{\displaystyle \frac{ d \langle \sigma_+a \rangle} {dt}=-i\delta
\langle \sigma_+a \rangle -\frac{\gamma+\kappa}{2} \langle \sigma_+a
\rangle +g \langle a^\dagger \sigma_z a \rangle + g \langle
\sigma_+\sigma_- \rangle}.
\end{array}
\end{equation}

The presence of the operator $\sigma_z$ is responsible for the
non-linear behavior of the two-level system. Nevertheless, for the
initial condition we consider (namely the state $\ket{e,0}$), the
dynamics of the atom-cavity system is restricted to the subspace
spanned by $\ket{e,0}, \ket{g,1}, \ket{g,0}$, where we have the
exact equalities $\langle \sigma_z a \rangle = -\langle a \rangle$
and $\langle  a^\dagger \sigma_z a \rangle = -\langle a^\dagger a
\rangle$. Note that the non-linearity induced by the operator
$\sigma_z$ exactly disappears. The behavior of a two-level system in
a field containing at most one photon cannot be distinguished from
the behavior of the two lower levels of a monomode cavity (or a
harmonic oscillator) indeed. This equivalence was exploited
in~\cite{bibi} to compute the linear transmission of a cavity
containing a two-level atom, and is the fundamental reason for the
simplicity of our method.

Equations~(\ref{equ:langevinq1}) are then identical to the equations
describing the evolution of the classical fields' amplitudes in the
two cavities ${\cal C}_1$ and ${\cal C}_2$ introduced above. In the
same way, equations~(\ref{equ:langevinq2}) correspond to the
equations governing the energies' evolutions in ${\cal C}_1$ and
${\cal C}_2$, with the initial conditions $|\alpha(0)|^2=1$ and
$|\beta(0)|^2=0$. As announced, the quantum regression theorem
allows us to conclude that the quantities $S_{at}(\omega)$ and
$S_{cav}(\omega)$ correspond to the spectra ${\cal S}_1(\omega)$ and
${\cal S}_2(\omega)$ emitted by ${\cal C}_1$ and ${\cal C}_2$ during
their relaxation process. Exact expressions for these spectra are
computed in appendix A. We obtain after normalization

\begin{equation}
\begin{array}{l}\label{equ:final}
{\displaystyle
S_{at}(\omega)=\frac{2((\kappa+\gamma)^2(4g^2+\kappa\gamma)+4\kappa\gamma\delta^2)
(\kappa^2+(\delta-2\omega)^2)}{\pi(4g^2(\kappa+\gamma)+\kappa((\kappa+\gamma)^2+4\delta^2)){\cal D}(\omega)}}\\
{\displaystyle
S_{cav}(\omega)=\frac{2((\kappa+\gamma)^2(4g^2+\kappa\gamma)+4\kappa\gamma\delta^2)}
{\pi(\kappa+\gamma){\cal D}(\omega)}},
\end{array}
\end{equation}

where we have introduced the quantity

\begin{equation}
{\cal D}(\omega)=16|\omega-\lambda_+|^2|\omega-\lambda_-|^2.
\end{equation}

Photons are emitted by the atom-cavity system with the complex
frequencies $\lambda_+$ and $\lambda_-$, where $\lambda_+$ and
$\lambda_-$ are the roots of the complex equation
$(\omega-\tilde{\omega}_1)(\omega-\tilde{\omega}_2)-g^2=0$, and are
studied in appendix B. If $\delta \rightarrow +\infty$, we have
$\lambda_+ \rightarrow \delta/2 - i\kappa/2$, which corresponds to
cavity-type photons, whereas $\lambda_-\rightarrow
-\delta/2-i\gamma/2$, which corresponds to atomic type photons. The
relative weight of each peak is given by a more detailed study of
the quantities $S_{cav}(\omega)$ and $S_{at}(\omega)$, which is done
in the following. We stress that equations~(\ref{equ:final})
generalize and match all previous results obtained on the SE of an
atom in a cavity~\cite{Carmichael,Agarwal,Rice,Alsing}.

It is interesting to relate the computed spectra to an expected
experimental signal. By connecting a time-resolved detector to each
channel of losses and summing both contributions, one has a density
of probability $p(t)$ to register a photon, fulfilling

\begin{equation}
p(t)=\gamma \langle \sigma_+\sigma_-(t) \rangle + \kappa \langle
a^\dagger a(t) \rangle.
\end{equation}

We have checked that

\begin{equation}
\gamma \int dt \langle \sigma_+\sigma_-(t) \rangle + \kappa \int dt
\langle a^\dagger a (t) \rangle =1,
\end{equation}

ensuring that one registers a photon with certainty. In the
following, we denote \\ $P_{at}= \gamma \int dt \langle
\sigma_+\sigma_-(t) \rangle$ and $P_{cav}=\kappa \int dt \langle
a^\dagger a (t) \rangle$ the probability for the quantum of energy
to escape in the atomic or in the cavity channel of losses
respectively.

If the detectors are frequency-resolved, one can define the total
density of probability per unit of frequency $p(\omega)$ to register
a photon, fulfilling

\begin{equation}
p(\omega)=P_{at}S_{at}(\omega)+P_{cav}S_{cav}(\omega).
\end{equation}

Before discussing the physical interpretation of the computed
quantities, we consider in the next section the links between the
spontaneous emission spectra and the fluorescence and absorption
spectra registered when the atom or the cavity is driven by a
classical field.

\section{Relation to fluorescence and absorption spectra}

\begin{figure}[h,t]
\begin{center}
\includegraphics[height=10cm]{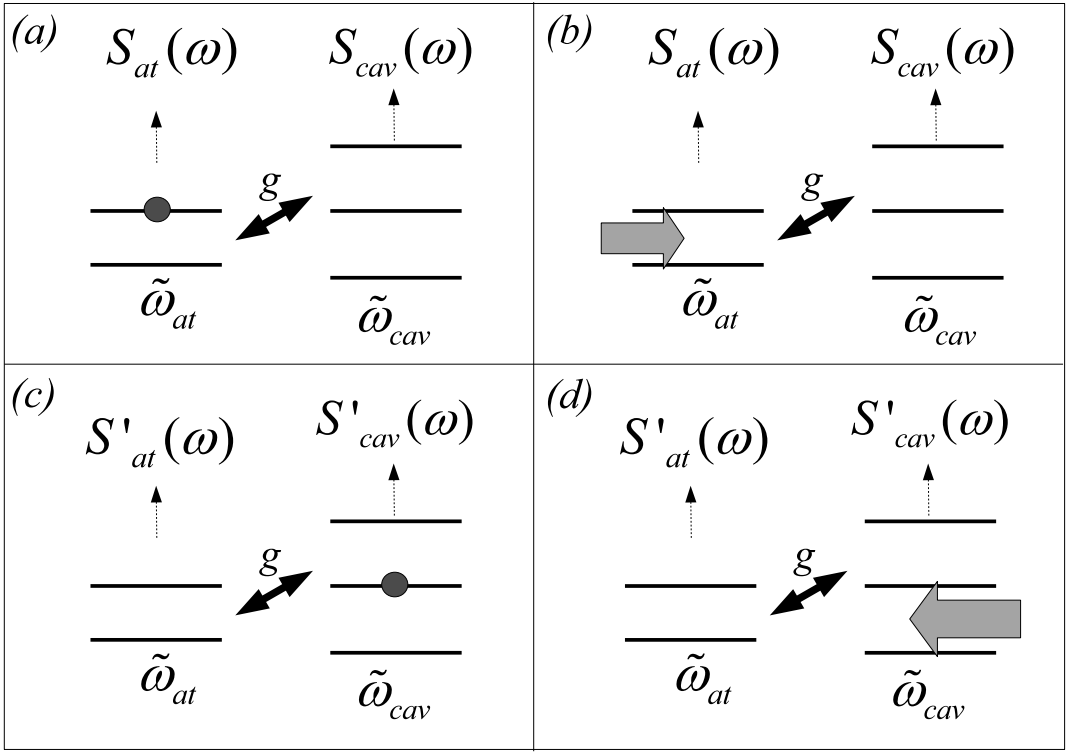}
\caption{\it A two-level atom of complex frequency
$\tilde{\omega}_{at}$ is coupled to a cavity of complex frequency
$\tilde{\omega}_{cav}$. $(a),(b)$ : Atomic-type spectroscopy. The
same spectra $S_{at}(\omega)$ and $S_{cav}(\omega)$ are recorded if
one initially excites the atom ($a$) or if one drives the atom with
a weak classical field ($b$). We have $S_{at}(\omega)=S_1(\omega)$,
$S_{cav}(\omega)=S_2(\omega)$ with
$\tilde{\omega}_1=\tilde{\omega}_{at}$ and
$\tilde{\omega}_2=\tilde{\omega}_{cav}$. $(c),(d)$ : Cavity-type
spectroscopy. The same spectra $S^{'}_{at}(\omega)$ and
$S^{'}_{cav}(\omega)$ are recorded if one initially feeds the cavity
with a single photon ($c$) or if one drives the cavity with a weak
classical field ($d$).  Now $S^{'}_{at}(\omega)=S_2(\omega)$,
$S^{'}_{cav}(\omega)=S_1(\omega)$ with
$\tilde{\omega}_1=\tilde{\omega}_{cav}$ and
$\tilde{\omega}_2=\tilde{\omega}_{at}$. $S_1(\omega)$ and
$S_2(\omega)$ are computed in appendix A. } \label{fig:symetrie}
\end{center}
\end{figure}

In this section, we compare the spectra spontaneously emitted by the
atom and the cavity when the atom is initially prepared in the
excited state, to the spectra obtained when the atom is driven by a
classical field of tunable frequency $\omega$ (atomic-type
spectroscopy). The solutions in the permanent regime are searched in
the frame rotating at the driving frequency, $\sigma_-(t)={\cal
\sigma}_\omega e^{-i\omega t}$, $a(t)=a_\omega e^{-i\omega t}$. We
pay attention to the case where the pump is too weak to saturate the
two-level system, so that the fluorescence spectrum emitted by the
atom is predominantly elastic. As a consequence, the spectra
respectively emitted by the atom and by the cavity are proportional
to $|\langle \sigma_\omega \rangle|^2$ and $|\langle a_\omega
\rangle |^2$. The supplementary coupling term in the hamiltonian
writes $i\hbar f(\sigma_+ e^{-i\omega t}-\sigma_- e^{i\omega t})$,
which modifies the Heisenberg equation~(\ref{equ:langevinq1}) for
the operator $\sigma_-$ in the following manner

\begin{equation}
\dot{\sigma}_-=\left(i\frac{\delta}{2}-\frac{\gamma}{2}\right)\sigma_-+g\sigma_za+g\sigma_zfe^{-i\omega
t}+F_s.
\end{equation}

Remembering that the system is mostly in the state $\ket{g,0}$, we
have $\langle \sigma_z \rangle \sim -1$, while we still have
$\langle \sigma_z a \rangle = -\langle a \rangle$. It is then
obvious that the mean values $\langle \sigma_\omega \rangle$ and
$\langle a_\omega \rangle$ follow the same evolution as the field's
amplitudes $\alpha_\omega$ and $\beta_\omega$ in the two cavities
${\cal C}_1$ and ${\cal C}_2$ introduced above, ${\cal C}_1$ being
now driven by a classical field $fe^{-i\omega t}$. This problem is
also studied in appendix A. In particular, it is shown that
$|\alpha_\omega|^2$ (resp. $|\beta_\omega|^2$) is proportional to
${\cal S}_1(\omega)$ (resp. ${\cal S}_2(\omega)$). As a consequence,
it is equivalent to initially excite the atom and record the spectra
emitted spontaneously by the atom $S_{at}(\omega)$ and by the cavity
$S_{cav}(\omega)$, or to drive the atom with a field of tunable
frequency and to observe the elastic fluorescence field emitted by
the atom and by the cavity. In the following these experiments will
be referred to as "atomic-type spectroscopy". They are schematized
in figures \ref{fig:symetrie}$a$ and \ref{fig:symetrie}$b$.

It is interesting to check that energy is conserved when the atom is
driven by a classical field. As it is shown in appendix A, energy
conservation in the permanent regime has the following form

\begin{equation}\label{equ:energie}
\gamma |\langle \sigma_\omega \rangle |^2+\kappa |\langle a_\omega
\rangle |^2=f (\langle \sigma_\omega \rangle + \langle \sigma_\omega
\rangle^*).
\end{equation}

The two terms on the left correspond to the atomic and cavity leaks,
the term on the right to the absorption of the system. These
quantities are expressed in number of photons per second.
Equation~(\ref{equ:energie}) can be rewritten in the following way,
evidencing the link between the absorption spectrum and the elastic
fluorescence spectra

\begin{equation}
f^2(P_{at}S_{at}(\omega)+P_{cav}S_{cav}(\omega))=f (\langle
\sigma_\omega \rangle + \langle \sigma_\omega \rangle^*).
\end{equation}

It is now obvious that $f^2 P_{at}S_{at}(\omega)$ (resp.
$f^2P_{cav}S_{cav}(\omega)$) represents the power scattered in the
atomic (resp. in the cavity) channel of losses when one drives the
atom with a classical field $fe^{-i\omega t}$.

This analysis is entirely valid if the parts of the cavity and of
the atom are exchanged, provided the pump's intensity is weak enough
not to saturate the two-level system. This condition is mandatory
indeed to make the atomic and the cavity's optical behavior
indistinguishable as it is explained in section II. In this case, it
is equivalent to drive the cavity mode and to register the field
respectively radiated by the atom $S^{'}_{at}(\omega)$ or by the
cavity $S^{'}_{cav}(\omega)$ (cavity-type spectroscopy), or to
initially feed the cavity with a single photon and to observe the
spectra spontaneously emitted by the atom or the cavity. Note that
this last experiment is quite delicate as it requires the ability to
prepare the cavity field in a Fock state~\cite{Patrice}. Now we have
$S^{'}_{cav}(\omega)=S_1(\omega)$ and
$S^{'}_{at}(\omega)=S_2(\omega)$, the complex frequencies having
been exchanged with respect to the previous case,
$\tilde{\omega}_1=\tilde{\omega}_{cav}+\delta/2-i\kappa/2$,
$\tilde{\omega}_2=\tilde{\omega}_{at}=-\delta/2-i\gamma/2$. In the
following these experiments will be referred to as "cavity-type
spectroscopy". They are schematized in figures \ref{fig:symetrie}$c$
and \ref{fig:symetrie}$d$.

Depending on the regime we consider, four experimental situations
are possible as summarized in table~\ref{tableau}.

\begin{equation}\label{tableau}
\begin{tabular}{|l|c|c|}
 \hline
 & Atomic-type spectroscopy & Cavity-type spectroscopy \\
 \hline
 Good emitter & A & B \\
  \hline
  Good cavity & C & D \\
   \hline
\end{tabular}
\end{equation}

\vspace{0.5cm}

Because of the symmetry between the atom and the cavity mentioned
above, the situations A and D (resp. B and C) are completely
equivalent. Namely, experiments A and D (resp. B and C) boil down to
registering the spectrum of a cavity when it is coupled to a broader
(resp. narrower) one. In section IV we discuss the spectrum emitted
in an atomic-type spectroscopy in the good emitter regime (A),
before turning to the physical interpretation of the spectrum
obtained in a cavity-type spectroscopy in the good cavity regime
(D). In section V, we consider the spectrum obtained in an
atomic-type spectroscopy operated in the good cavity regime (C),
before considering the symmetrical experiment of a cavity-type
spectroscopy in the good emitter regime (B).

\section{Atomic-type spectroscopy in the good emitter regime}

\begin{figure}[h,t]
\begin{center}
\includegraphics[height=10cm]{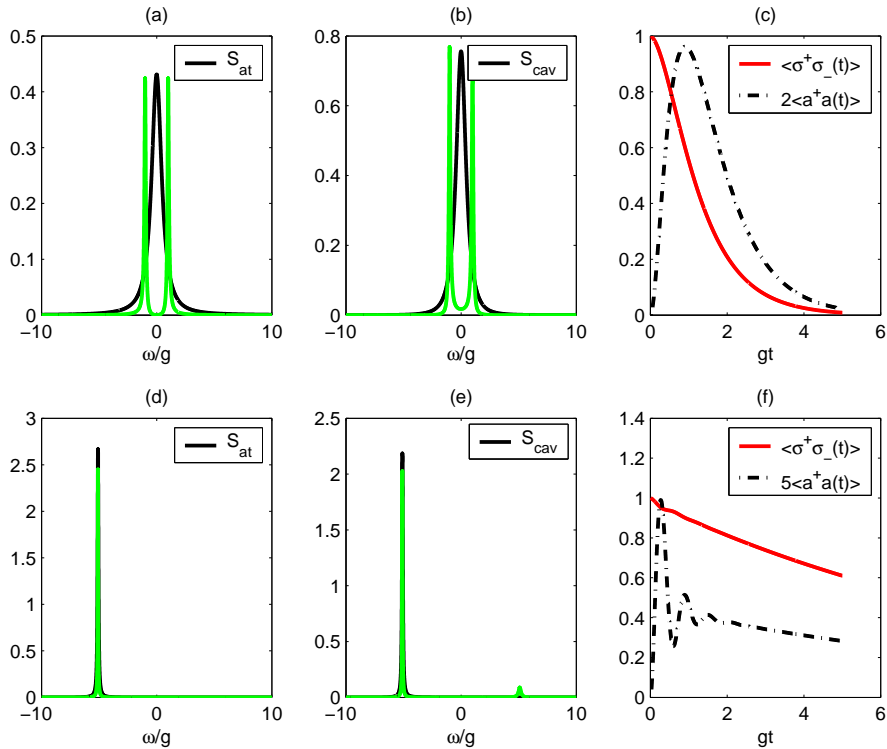}
\caption{\it Spectra emitted by the atom $(a)$ and by the cavity
$(b)$ during an atomic-type spectroscopy in the good-emitter regime
(resonant case). We took $\gamma=0.05,g=1$. Black line : $\kappa=5$
(weak coupling regime). Green line : $\kappa=0.25$ (strong coupling
regime). $(c)$ Evolution of the atomic (red solid line) and cavity
(black dashed-dotted line) populations as a function of time in the
weak coupling case in a SE experiment. $(d)$, $(e)$ and $(f)$ : same
study if the atom and the cavity are detuned. We took $\delta=10$.}
\label{fig:bad-cavity}
\end{center}
\end{figure}

\subsection{Resonant case}

To begin with, we study the spectra $S_{at}(\omega)$ and
$S_{cav}(\omega)$ emitted during an atomic-type spectroscopy in the
good emitter regime, in the resonant case. As underlined above, the
good emitter regime was satisfied in most experiments until now. The
limit of this case consists in a perfect atom coupled to a finite
bandwidth cavity, which was extensively studied in~\cite{Agarwal}
and is well understood. We just recall the main results for sake of
completeness.

We have represented in figures~\ref{fig:bad-cavity}$a$ and
\ref{fig:bad-cavity}$b$ the spectra $S_{at}$ and $S_{cav}$ in the
strong coupling case (dotted line) and in the weak coupling case
(solid line). We recover that in the strong coupling case, the
spectra consist in the vacuum Rabi doublet evidenced
in~\cite{Sanchez,Carmichael}, whereas in the weak coupling case,
they reduce to a single peak. We stress that $S_{at}$ and $S_{cav}$
are identical. In the weak coupling case, the cavity mode can be
adiabatically eliminated indeed, so that both oscillators undergo
the same dynamics, leading to identical emission properties. This
analysis is confirmed when one looks at
figure~\ref{fig:bad-cavity}$c$, where the evolution of the atomic
and cavity populations is represented as a function of time. After
the transient, the cavity and the atom follow the same evolution. In
this picture, the cavity behaves as a relaxation channel for the
atom, the coupling rate to this channel scaling like $g^2/\kappa$,
as pointed out in~\cite{Agarwal}. In the case of a two-level system
with very few losses ($\gamma \ll g^2/\kappa$), the coupling to the
resonant cavity leads to an enhancement of the SE rate (Purcell
effect~\cite{Purcell}).

\subsection{Non-resonant case}

We have represented in figures~\ref{fig:bad-cavity}$d$ and
\ref{fig:bad-cavity}$e$ the spectra $S_{at}$ and $S_{cav}$ in the
strong coupling case (dotted line) and in the weak coupling case
(solid line) when the atom and the cavity are detuned. As it can be
seen, both the atom and the cavity emit photons at the atomic
frequency. The previous explanation is still valid, the cavity mode
following adiabatically the atomic mode and behaving like a
relaxation channel for the atom. Nevertheless, its efficiency is
strongly reduced with respect to the resonant case, the coupling
rate being now $g^2 \kappa/\delta^2$. Note that these results can be
obtained in the perturbative regime, by studying the spontaneous
emission of a two-level system in a structured continuum with the
Fermi-golden rule as underlined by~\cite{Andreani}.

\subsection{Symmetrical situation}

Here we analyze the physical meaning of the symmetrical experiment,
namely a cavity-type spectroscopy in the good-cavity regime. As
explained in section III, such an experiment can be realized by
driving the cavity with a classical field of tunable frequency and
registering the fluorescence spectra emitted by the atom and by the
cavity, or by feeding the cavity with a single photon and recording
the spontaneous emission spectra. In the frame of solid-state
physics, both experiments are quite delicate, the first one
requiring the ability to filtrate the pump's light, the second one
the capacity to prepare the cavity in a non-classical state. The
equivalent of the Purcell effect would be the enlargement of the
cavity linewidth due to its coupling to the lossy atom. This
enlargement is simply due to the fact that the field in the cavity
mode can efficiently be scattered in the atomic channel of losses.

\section{Atomic-type spectroscopy in the good cavity regime}

\begin{figure}[h,t]
\begin{center}
\includegraphics[height=10cm]{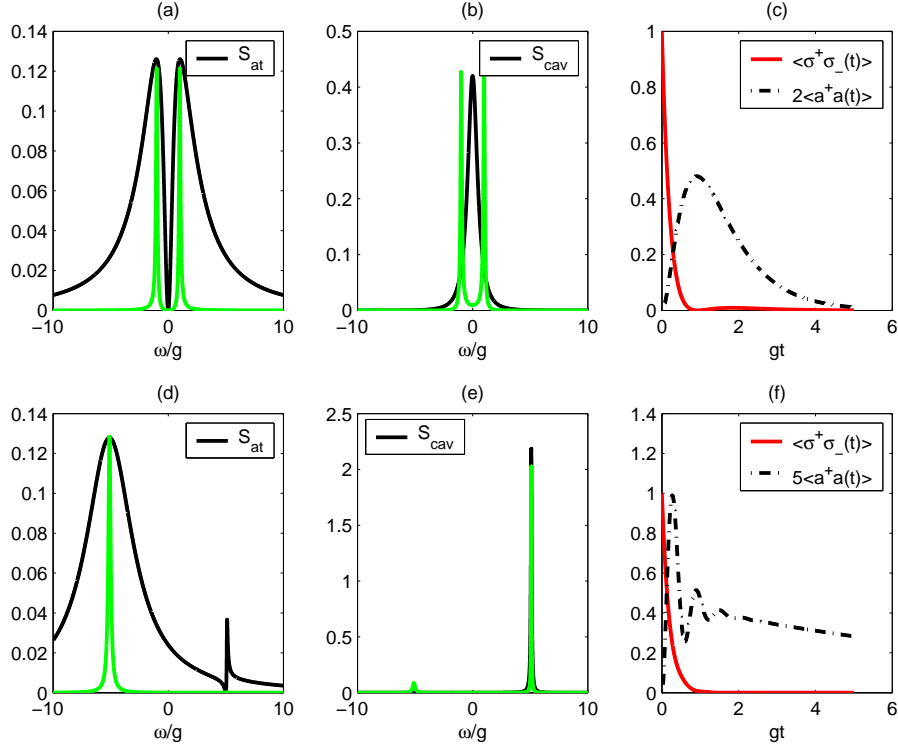}
\caption{\it Spectra emitted by the atom $(a)$ and by the cavity
$(b)$ during an atomic-type spectroscopy in the good-cavity regime
(resonant case). We took $\kappa=0.05,g=1$. Black line : $\gamma=5$
(weak coupling regime). Green line : $\gamma=0.25$ (strong coupling
regime). $(c)$ Evolution of the atomic (red solid line) and cavity
(black dashed-dotted line) populations as a function of time in the
weak coupling case in a SE experiment. $(d)$, $(e)$ and $(f)$ : same
study in the case where the atom and the cavity are detuned. We took
$\delta=10$.}\label{fig:good-cavity}
\end{center}
\end{figure}

\subsection{Resonant case}
We come to the interpretation of an atomic-type spectroscopy
experiment held in the good-cavity regime, $\gamma
> \kappa$. As it is shown in introduction, this regime is reached is
an ever increasing number of experiments belonging to a wide range
of systems, thanks to recent progress in the production of high Q
cavities. We focus on the resonant case first. We have represented
in figure~\ref{fig:good-cavity}$a$ and \ref{fig:good-cavity}$b$ the
spectra emitted by the atom and by the cavity in the strong coupling
regime (dotted line) and in the weak coupling regime (solid line).
As previously, in the strong coupling regime, both spectra consist
in the vacuum Rabi doublet. On the contrary, in the weak coupling
regime, the spectra respectively emitted by the atom and the cavity
are dramatically different. As it can be seen in
figure~\ref{fig:good-cavity}$a$, the presence of a very high Q
resonator induces a hole in the atomic spectrum, which was already
pointed out by~\cite{Alsing,Rice}. In these references, the hole is
attributed to a destructive interference between the driving field
and the intra-cavity field, so that the atom decouples and doesn't
fluoresce anymore. This phenomenon was referred to as Cavity Induced
Transparency (CIT). In a classical picture, it can be attributed to
a double resonance effect, happening if one drives a broad cavity,
coupled to a narrow cavity. Our computation of the atomic and cavity
populations as functions of time, which is represented in
figure~\ref{fig:good-cavity}$c$, offers a new insight on CIT in the
context of a spontaneous emission experiment. Contrary to what
happens in the bad cavity case, the cavity evolves much slower than
the atom, trapping the fraction of photon that is resonant with it,
before re-emitting it in its own channel of losses. The spectrum
emitted by the cavity $S_{cav}(\omega)$ is a peak of typical width
$4g^2/\gamma$, which is much lower than $\gamma$. The cavity behaves
thus as a very narrow spectral filter.

\subsection{Non-resonant case}
We come to the case where the atom and the cavity are detuned. As
before, the analysis presented in this paragraph is valid in the
strong and the weak coupling regime. One can see in
figure~\ref{fig:good-cavity}$d$ that the atom still emits photons at
its own frequency. On the contrary, as it appears in
figure~\ref{fig:good-cavity}$e$, $S_{cav}(\omega)$ is a peak
centered around the cavity frequency, which is totally different
from what happens in the good emitter regime. Like in the resonant
case, the cavity traps the fraction of photon resonant with it and
reemit it in its own channel of losses with its own characteristics
(frequency and linewidth). The absorption of the atom-cavity system
at the cavity frequency is usually attributed to Rayleigh
scattering, which is the non resonant scattering of photons by the
atom in the cavity mode, enhanced by the presence of the
mirrors~\cite{serge}. In a spontaneous emission picture, this can be
seen as a simple dressed-state effect. The initial state of the
system is $\ket{e,0}$ indeed, which has an overlap with the
"cavity-type" dressed state scaling like $g^2/\delta^2$, accounting
for the emission of photons at this wavelength~\cite{serge}. Our
results go beyond this analysis, as they allow to distinguish what
is emitted by the cavity from what is emitted by the atom, and show
that cavity-type photons are emitted in the cavity channel of
losses, whereas atomic-type photons are emitted in the atomic one.
By geometrically coupling a detector predominantly to the cavity
channel of losses, one can select cavity-type rather than
atomic-type photons. This feature was experimentally observed for
quantum dots strongly coupled to semiconducting
cavities~\cite{Hennessy,Press}, where a significant fraction of
photons were emitted at the cavity frequency. For a more detailed
analysis of these results, it will be necessary to take into account
the impact of additional sources of excitonic decoherence in
experimental systems.

\subsection{Symmetrical experiment}
We come to the discussion of the reversed experiment, namely the
cavity-type spectroscopy in the good emitter regime. A hole in the
resonance spectrum of a cavity coupled to a narrow atom has been
predicted by~\cite{Rice}. It was also evidenced in~\cite{bibi}, in
the particular case of a one-dimensional cavity, and referred to as
Dipole Induced Reflexion. Moreover, it is shown in~\cite{bibi} that
the medium consisting in a two-level atom coupled to a directional
cavity provides a giant optical non-linear medium sensitive at the
single-photon level. This feature was recently observed in the case
of a photonic crystal cavity strongly coupled to a single quantum
dot~\cite{Englund}.

\section{Conclusion}
We have shown that the SE of an atom in a cavity has a classical
counterpart allowing to compute exact expressions for the spectra
emitted by the atom and the cavity, in the most general case where
the systems have arbitrary linewidths and frequencies. The basic
reason for the existence of this classical equivalent is the
vanishing of the atomic non-linearity in the relevant subspace
$\ket{e,0}$, $\ket{g,1}$, $\ket{g,0}$. This is due to the fact that
when it is weakly excited, the two states of a two-level system
cannot be distinguished from the two lower states of a monomode
cavity. We have evidenced the equivalence between the spontaneous
emission and elastic fluorescence spectra. Our study provides all
previous results with a common theoretical frame, and goes beyond as
it allows to draw a comparison between the atomic and cavity
spectra. The comparison is particularly striking for a detuned
atom-cavity system. The spectra are identical in the good emitter
regime, and dramatically different in the good cavity regime. In the
first case, the cavity behaves as a relaxation channel. On the
contrary, in the second case, the cavity mode can be considered as a
narrow filter, trapping and reemitting a fraction of photons with
its own characteristics (frequency and linewidth) in its own channel
of losses. If the radiation patterns of the atomic leaky modes and
of the cavity are distinguishable, it is possible to selectively
detect photons at the cavity frequency. Besides its conceptual
interest, this good-cavity regime could be exploited in original
devices, such as spectrally-tunable single photon sources based on
cavity tuning.

\section{Acknowledgment}
This work was supported by the Agence Nationale de la Recherche
under the project IQ-Nona.

\appendix

\section{Classical coupled cavities}

In this appendix we compare the relaxation and the fluorescence
spectra emitted by two coupled cavities ${\cal C}_1$ and ${\cal
C}_2$. We denote $\alpha$ and $\beta$ the complex amplitudes of the
field in each cavity, $\tilde{\omega}_1$ and $\tilde{\omega}_2$
their respective complex frequencies, and $g$ the coupling strength
between the two cavities. We define the Fourier transform
$\tilde{\psi}(\omega)$ of a function $\psi(t)$ in the following way

\begin{equation}
\tilde{\psi}(\omega)=\frac{1}{2\pi}\int dt \psi(t) e^{i\omega t}
\end{equation}

First we consider the relaxation of the total system when the first
cavity is fed with a classical field at time $t=0$. The evolution of
the system is described by the set of equations

\begin{equation}\label{equ:dyn}
\begin{array}{l}
\dot{\alpha}+i\tilde{\omega}_1 \alpha+g \beta= 0 \\
\dot{\beta}+i\tilde{\omega}_2 \beta - g \alpha=0,
\end{array}
\end{equation}

the initial conditions being $\alpha(0)=1$ and $\beta(0)=0$. The
solutions are linear combinations of exponential functions at the
eigenfrequencies $\lambda_+$ and $\lambda_-$, where $\lambda_+$ and
$\lambda_-$ are the roots of the equation
$(\lambda-\tilde{\omega}_1)(\lambda-\tilde{\omega}_2)-g^2=0$. Their
form is studied in appendix B.

Taking into account the initial conditions we find

\begin{equation}\label{equ:dyn}
\begin{array}{l}
{\displaystyle
\alpha(t)=\frac{\lambda_+-\tilde{\omega}_2}{\lambda_+-\lambda_-}e^{-i\lambda_+
t}-\frac{\lambda_--\tilde{\omega}_2}{\lambda_+-\lambda_-}e^{-i\lambda_-
t}}\\
{\displaystyle
\beta(t)=\frac{ig}{\lambda_+-\lambda_-}\left(e^{-i\lambda_+
t}-e^{-i\lambda_- t}\right)}.
\end{array}
\end{equation}

One denotes $S_1(\omega)$ and $S_2(\omega)$ the spectra emitted by
each cavity during the relaxation process, fulfilling
(Wiener-Khinchine theorem)

\begin{equation}
\begin{array}{l}
{\displaystyle S_1(\omega)=\frac{|\tilde{\alpha}(\omega)|^2}{\int
d\omega |\tilde{\alpha}(\omega)|^2}=\frac{\int dt d\tau
\alpha^*(t+\tau) \alpha(t)}{2\pi \int dt
|\alpha(t)|^2}} \\
{\displaystyle S_2(\omega)=\frac{|\tilde{\beta}(\omega)|^2}{\int
d\omega |\tilde{\beta}(\omega)|^2}=\frac{\int dt d\tau
\beta^*(t+\tau) \beta(t)}{2\pi \int dt |\beta(t)|^2}}
\end{array}
\end{equation}

From equations~(\ref{equ:dyn}), one easily finds

\begin{equation}\label{equ:interm0}
\begin{array}{l}
{\displaystyle
\tilde{\alpha}(\omega)=\frac{-i}{\lambda_+-\lambda_-}\left[\frac{\lambda_+-\tilde{\omega}_2}{\lambda_+-\omega}-
\frac{\lambda_--\tilde{\omega}_2}{\lambda_--\omega}\right]}\\
{\displaystyle
\tilde{\beta}(\omega)=\frac{g}{\lambda_+-\lambda_-}\left[\frac{1}{\lambda_+-\omega}-
\frac{1}{\lambda_--\omega}\right]}.
\end{array}
\end{equation}

which allows us to get, up to a normalization factor,

\begin{equation}
\begin{array}{l}
{\displaystyle
S_1(\omega)\equiv\frac{1}{|\lambda_+-\lambda_-|^2}\left|\frac{\lambda_+-\tilde{\omega}_2}{\lambda_+-\omega}-
\frac{\lambda_--\tilde{\omega}_2}{\lambda_--\omega}\right|^2}\\
{\displaystyle
S_2(\omega)\equiv\frac{1}{|\lambda_+-\lambda_-|^2}\left|\frac{1}{\lambda_+-\omega}-
\frac{1}{\lambda_--\omega}\right|^2}.
\end{array}
\end{equation}

We focus now on the case where the first cavity is driven at the
frequency $\omega$. The set of equations describing the dynamics
of the system writes now

\begin{equation}\label{equ:drive}
\begin{array}{l}
\dot{\alpha}+i\tilde{\omega}_1 \alpha+g \beta= f e^{-i\omega t} \\
\dot{\beta}+i\tilde{\omega}_2 \beta - g \alpha=0.
\end{array}
\end{equation}

$f$ is taken real. The solutions in the permanent regime have the form
$\alpha(t)=\alpha_\omega e^{-i\omega t}$ and
$\beta(t)=\beta_\omega e^{-i\omega t}$. Recalling the equality
$(\omega-\tilde{\omega}_1)(\omega-\tilde{\omega}_2)-g^2=(\omega-\lambda_+)(\omega-\lambda_2)$,
we immediately have

\begin{equation}\label{equ:interm2}
\begin{array}{l}
{\displaystyle \alpha_\omega=
\frac{i(\omega-\tilde{\omega}_2)f}{(\omega-\lambda_+)(\omega-\lambda_-)}}\\
{\displaystyle \beta_\omega=
\frac{-gf}{(\omega-\lambda_+)(\omega-\lambda_-)}}.
\end{array}
\end{equation}

\begin{figure}[h,t]
\begin{center}
\includegraphics[height=10cm]{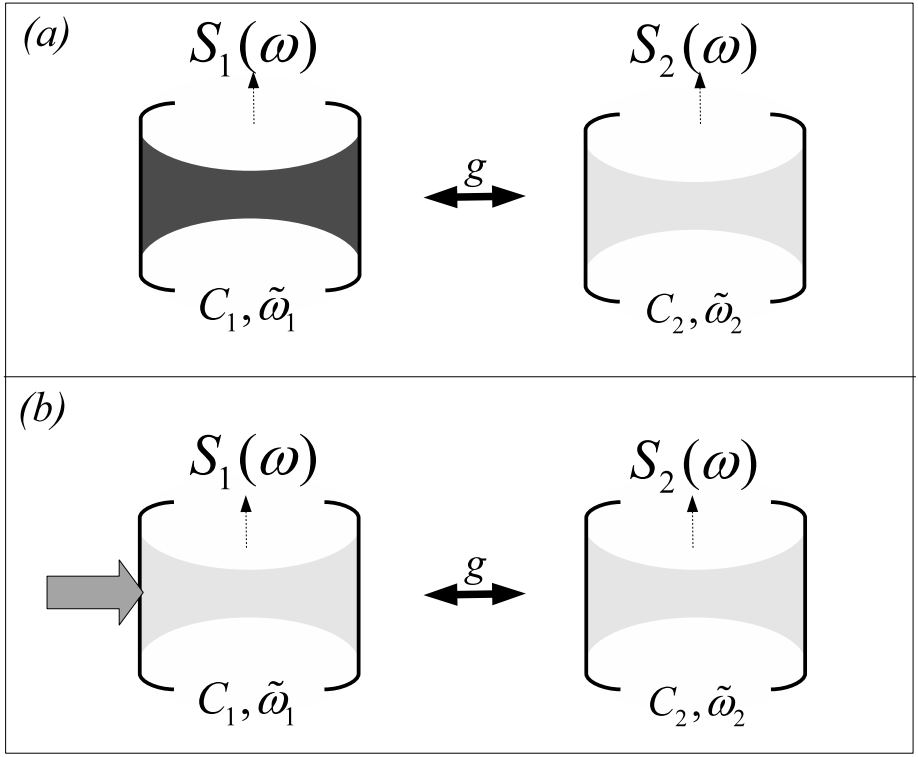}
\caption{\it Two cavities ${\cal C}_1$ and ${\cal C}_2$ of
respective complex frequencies $\tilde{\omega}_1$ and
$\tilde{\omega}_2$ are coupled with a strength $g$. ${\cal C}_1$ (resp. ${\cal C}_2$)
emits the spectrum $S_1(\omega)$ (resp. $S_2(\omega)$), whether
${\cal C}_1$ is initially fed with a classical field ($a$), or if it
is driven with a classical field of tunable frequency ($b$).}
\label{fig:symetrie-cav}
\end{center}
\end{figure}

Note that equations~(\ref{equ:interm0}) can be rewritten

\begin{equation}\label{equ:interm}
\begin{array}{l}
{\displaystyle
\tilde{\alpha}(\omega)=\frac{i(\tilde{\omega_2}-\omega)}{(\lambda_+-\omega)(\lambda_--\omega)}}\\
{\displaystyle
\tilde{\beta}(\omega)=\frac{-g}{(\lambda_+-\omega)(\lambda_--\omega)}},
\end{array}
\end{equation}

showing that $\tilde{\alpha}(\omega)\propto\alpha_\omega$ and
$\tilde{\beta}(\omega)\propto\beta_\omega$. The spectra emitted by
each cavity being respectively proportional to $|\alpha_\omega|^2$
and $|\beta_\omega|^2$, they also correspond to ${\cal S}_1(\omega)$
and ${\cal S}_2(\omega)$. As a consequence, it is equivalent to
initially feed the first cavity with a classical field and to
observe the spectra emitted by each cavity ${\cal S}_1(\omega)$ and
${\cal S}_2(\omega)$ during the relaxation process, or to drive the
first cavity with a classical field and to observe the fluorescence
spectra radiated by each cavity. The corresponding experiments are
schematized in figure~\ref{fig:symetrie-cav}.

Finally it is interesting to check energy conservation of the system. In the permanent regime its energy is constant and
equals $E_\omega=|\alpha_\omega|^2+|\beta_\omega|^2$, where we have expressed $E_\omega$ in number of photons. We have

\begin{equation}
\frac{dE_\omega}{dt}=0=i(\tilde{\omega}_1^*-\tilde{\omega}_1)|\alpha_\omega|^2+
i(\tilde{\omega}_2^*-\tilde{\omega}_2)|\beta_\omega|^2+
f(\alpha_\omega+\alpha_\omega^*).
\end{equation}

The first two terms represent the leaks of each cavity. The third term exactly balances the leaks and
corresponds to the absorption of the system.

\section{Eigenfrequencies}

In this appendix we study the eigenfrequencies $\lambda_+,\lambda_-$
of the coupled cavities system. They fulfil

\begin{equation}
\begin{array}{l}
{\displaystyle \lambda_++\lambda_-=\tilde{\omega}_1+\tilde{\omega}_2} \\
{\displaystyle
\lambda_+\lambda_-=\tilde{\omega}_1\tilde{\omega}_2-g^2}
\end{array}
\end{equation}

In the resonant case, we have
$\lambda_\pm=i\frac{\kappa+\gamma}{4}\pm\sqrt{g^2-\left(\frac{\kappa-\gamma}{4}\right)^2}$.
The system has two distinct oscillation frequencies if $4g >
|\kappa-\gamma|$, as pointed out by Andreani and
coworkers~\cite{Andreani}. This is the case in the strong coupling
regime defined as $g \gg \kappa, \gamma$, where the eigenfrequencies
of the system are simply $\pm g$. In the general case the solutions
can be written ${\displaystyle \lambda_\pm=\pm\frac{a}{2} - i
\frac{b_\pm}{2}}$, with

\begin{equation}
\begin{array}{l}
a^2={\displaystyle
2g^2+\frac{\delta^2}{2}-\frac{1}{2}\left(\frac{\kappa-\gamma}{2}\right)^2+
\sqrt{\left(2g^2-\frac{1}{2}\left(\frac{\kappa-\gamma}{2}\right)^2\right)^2
+ \delta^2
\left(2g^2-\frac{1}{2}\left(\frac{\kappa-\gamma}{2}\right)^2\right) + \left(\frac{\delta^2}{2}\right)^2 } }\\
{\displaystyle b_+=\frac{\kappa+\delta}{2}+\frac{\delta}{2a}(\kappa-\gamma)} \\
{\displaystyle b_-=\frac{\kappa+\delta}{2}-\frac{\delta}{2a}(\kappa-\gamma)} \\
\end{array}
\end{equation}

If the detuning is strong $\delta \gg g, \kappa, \gamma$, we have
$\lambda_+\sim \delta/2 -i\kappa/2$, $\lambda_-\sim -\delta/2
-i\gamma/2$ and we recover the complex frequency of the undressed
cavities. It is interesting to consider the limits of these
expressions in the strong coupling case where $g \gg \kappa,
\gamma$. After development the quantity $a$ takes the following form

\begin{equation}
a^2=\delta^2+4g^2-\left(\frac{\kappa-\gamma}{2}\right)^2
\sin^2(2\theta)
\end{equation}

where we have introduced the mixing angle $\theta$, checking $\tan
(2\theta) = 2g/\delta$. We obtain after some little algebra
$b_+=\kappa \cos^2 \theta + \gamma \sin ^2 \theta$ and $b_-=\kappa
\sin^2 \theta + \gamma \cos ^2 \theta$. In the strong coupling
regime, the poles $\lambda_+$ and $\lambda_-$ correspond to the
complex frequencies of the first manifold's dressed states of the
atom-cavity system as computed in~\cite{serge}.

\end{document}